\documentclass[]{spie}  

\usepackage{amsmath,amsfonts,amssymb}
\usepackage{graphicx}
\usepackage[colorlinks=true, allcolors=blue]{hyperref}
\usepackage{multirow}
\usepackage{array}
\usepackage{longtable}
\usepackage{soul}
\usepackage{enumitem}

\usepackage{xcolor,colortbl}
\usepackage[]{aas_macros}

\graphicspath{{figures/}}

\title{High-Contrast Imaging at First-Light of the GMT: The Preliminary Design of GMagAO-X}

\author[a]{Jared R. Males}
\author[a]{Laird M. Close}
\author[a,b]{Sebastiaan Y. Haffert}
\author[a,c]{Maggie Y. Kautz}
\author[a]{Doug Kelly}
\author[a]{Adam Fletcher}
\author[a]{Thomas Salanski}
\author[a]{Olivier Durney}
\author[a]{Jamison Noenickx}
\author[a]{John Ford}
\author[a]{Victor Gasho}
\author[a]{Logan Pearce}
\author[a,c]{Jay Kueny}
\author[a,c,d,e]{Olivier Guyon}
\author[f]{Alycia Weinberger}
\author[g]{Brendan Bowler}
\author[g]{Adam Kraus}
\author[h]{Natasha Batalha}

\affil[a]{Steward Observatory, University of Arizona}
\affil[b]{Leiden Observatory, Leiden University, The Netherlands}
\affil[c]{James C. Wyant College of Optical Sciences, University of Arizona, USA}
\affil[d]{Subaru Telescope, National Astronomical Observatory of Japan}
\affil[e]{Astrobiology Center, National Institutes of Natural Sciences, Japan}
\affil[f]{Carnegie Institution for Science, Washington, D.C., USA}
\affil[g]{University of Texas at Austin, USA}
\affil[h]{NASA Ames Research Center, USA}

\authorinfo{Further author information: (Send correspondence to J.R.M.)\\J.R.M.: E-mail: jrmales@arizona.edu}
\begin{document} 

\maketitle

\begin{abstract}
We present the preliminary design of GMagAO-X, the first-light high-contrast imager planned for the Giant Magellan Telescope.   GMagAO-X will realize the revolutionary increase in spatial resolution and sensitivity
provided by the 25 m GMT.  It will enable, for the first time, the spectroscopic characterization of nearby potentially habitable terrestrial exoplanets orbiting late-type stars.   Additional science cases include:
reflected light characterization of mature giant planets; measurement of young extrasolar giant planet variability; characterization of circumstellar disks at unprecedented spatial resolution; characterization of
benchmark stellar atmospheres at high spectral resolution; and mapping of resolved objects such as giant stars and asteroids.  These, and many more, science cases will be enabled by a 21,000 actuator extreme adaptive
optics system, a coronagraphic wavefront control system, and a suite of imagers and spectrographs.  We will review the science-driven performance requirements for GMagAO-X, which include achieving a Strehl ratio of
70\% at 800 nm on 8th mag and brighter stars, and post-processed characterization at astrophysical flux-ratios of 1e-7 at 4 lambda/D (26 mas at 800 nm) separation.  We will provide an overview of the resulting
mechanical, optical, and software designs optimized to deliver this performance.  We will also discuss the interfaces to the GMT itself, and the concept of operations.   We will present an overview of our end-to-end
performance modeling and simulations, including the control of segment phasing, as well as an overview of prototype lab demonstrations.   Finally, we will review the results of Preliminary Design Review held in
February, 2024.
\end{abstract}


\section{INTRODUCTION}
\label{sec:intro}  

The 25-40 m Extremely Large Telescopes (ELTs) will transform the study of extrasolar planets.  Their larger collecting area, compared to current generation 5-10 m telescopes, will dramatically improve the sensitivity of exoplanet studies including radial velocity measurements and transit spectroscopy on nearby planets.  The gains for resolved direct imaging of exoplanets will be even more dramatic.  The most profound result of this transformative leap in capabilities will be the ability to search for the signatures of life on planets orbiting other stars.  The Astro2020 Decadal Survey\cite{2021pdaa.book.....N} placed the search for life on exoplanets as a Priority Area for astrophysics over the next decade. In ``Pathways to Habitable Worlds'', Astro2020 found that one of the four key capabilities needed to achieve this is: \textbf{\textit{Ground-based extremely large telescopes equipped with high-resolution spectroscopy, high-performance adaptive optics, and high-contrast imaging.}}\cite{2021pdaa.book.....N}

Preparing the instrumentation suite of the ELTs to deliver on this promise is crucial, but also a demanding technological challenge.  It is well known that in the background-noise-limited case, the sensitivity (time to signal-to-noise, SNR) to point sources of diffraction-limited telescopes scales with diameter $D^4$.  Just as importantly, the improvement in spatial resolution with $1/D$ will allow direct imaging at smaller separations than currently possible with today's telescopes.  However, \textit{\textbf{the application of these scaling laws to direct imaging is non-trivial}}.  $D^4$ only obtains if the noise scales with $D^2$ and exposure time, but this is well-known to not be the case at small separations for typical high-contrast imaging instruments.  This is due to the internal systematics driven quasi-static speckle problem\cite{2009ARA&A..47..253O}.  Only if quasi-static speckles are controlled and suppressed such that residual speckle lifetimes are short will such an instrument achieve $D^4$\cite{2021PASP..133j4504M}.  Furthermore, the science foreseen by Astro2020 requires working at small separations, $\lesssim 5 \lambda/D$,  with coronagraphs.  This requires exquisite wavefront sensing and control (WFS\&C) and so places demanding requirements on a direct imaging instrument.

The 25 m Giant Magellan Telescope (GMT) presents the best opportunity to realize the vision of Astro2020 early in the ELT era.  Given its unique aperture and smaller $D$, an instrument can be constructed to achieve higher Strehl with fewer deformable mirror (DM) actuators.  This  makes the GMT the ideal facility for short wavelength Extreme AO (ExAO) \cite{2014SPIE.9148E..1MC,2018ARA&A..56..315G}. The large-segment design of the GMT provides a straightforward way to deploy mature, well tested DM technology with high actuator density.   Equipped with an ExAO-fed coronagraph, the GMT will characterize large numbers of \textit{temperate}, \textit{mature} exoplanets for the first time\cite{2012SPIE.8447E..1XG,2014SPIE.9148E..20M}, including terrestrial \textit{potentially habitable} exoplanets.

Here we present the results of the preliminary design for GMagAO-X, an instrument in development to provide a first-light ExAO and coronagraph facility for GMT.  We will review the science cases, science-driven requirements, and the mechanical, optical, electrical, and software \& control preliminary designs. GMagAO-X underwent a successful preliminary design review (PDR) in February, 2024 and is now preparing for final design.

\begin{table}
\centering
\footnotesize
\caption{Key Instrument Parameters of GMagAO-X \label{tab:params}}
\begin{tabular}{|l|c|c|c|l|}
\hline
Parameter   &    Requirement    &    Goal  & Stretch Goal   &  Notes \\
\hline
\hline
Wavelength Coverage & 600 -- 1900 nm & 450 -- 1900 nm & 350--1900 nm & \multirow{2}{3.5cm}{GMAGX-SCI-002, GMAGX-SCI-006} \\
&&&&\\
\hline
Spatial Resolution  & 4.9 mas & 3.7 mas & 2.8 mas & GMAGX-SCI-002 \\
\hline
\multirow{3}{*}{Spectral Resolution} & BB: 10\% bands  &        & & \multirow{3}{3.5cm}{GMAGX-SCI-003, GMAGX-SCI-004, GMAGX-SCI-005}\\
                                    &  Low: 20--100             &          & & \\
                                    &  High: 1000--65,000        & 100,000  & & \\
\hline
Guide Star I Magnitude Range & -1.5 -- 13 & -1.5 -- 15  &  & GMAGX-SCI-007 \\
\hline
Field of View  & 3'' x 3'' & 3.5'' x 6'' &  & GMAGX-SCI-010 \\
\hline
\multirow{3}{*}{Coronagraph Contrast}  & \multirow{3}{*}{1e-7 @ 4$\lambda/D$} & 1e-7 @ 2$\lambda/D$ &  1e-8 @ 1$\lambda/D$  & \multirow{3}{3.5cm}{5$\sigma$ planet:star flux. GMAGX-SCI-001, GMAGX-SCI-008}\\
                                       &                     & 1e-8 @ 6$\lambda/D$ &  1e-9 @ 5$\lambda/D$  & \\
                      &&&&\\
\hline

\end{tabular}
\end{table}

\section{SCIENCE CASES AND REQUIREMENTS}

The GMagAO-X science team produced the following core science cases.  We briefly state them here, see the \textit{GMagAO-X Instrument Science Requirements Document} (GMAGX-DOC-0001) for detailed discussion.

\begin{enumerate}
\itemsep -5pt
    \item Reflected Light Imaging and the Search for New Life
    \vspace{-3pt}
        \begin{enumerate}
        \itemsep -1pt
            \item Characterizing Temperate Giant Exoplanets
            \item Terrestrial Planets: Detecting and Characterizing the Atmosphere
                \begin{enumerate}
                    \item Detection of biosignatures (e.g. $H_2 O$, $O_2$).
                \end{enumerate}
            \item Reconnaissance for Habitable World's Observatory
        \end{enumerate}

    \item Characterizing Young Extrasolar Giant Planets (EGPs)
    \vspace{-3pt}
        \begin{enumerate}
        \itemsep -1pt
            \item Young Self-Luminous EGPs
            \item Rotational Periods of EGPs
            \item Cloud 3-D Structure
        \end{enumerate}

    \item Planet Formation at Low Mass and Small Separation
    \vspace{-3pt}
        \begin{enumerate}
        \itemsep -1pt
            \item Accretion onto Planets and Disks
        \end{enumerate}

    \item Circumstellar Disk Structure and Disk-Planet Interactions

    \item Stellar Evolution
    \vspace{-3pt}
        \begin{enumerate}
        \itemsep -1pt
            \item Benchmark Binary System Characterization
            \item White Dwarf / Main Sequence Star Binaries
                \begin{enumerate}
                    \item White Dwarf Pollution
                \end{enumerate}
            \item Interacting Stars
            \item Resolved Stars
        \end{enumerate}

    \item Solar System Science
    \vspace{-3pt}
        \begin{enumerate}
        \itemsep -1pt
            \item Asteroid Density Measurements
            \item Solar System Volatiles
        \end{enumerate}

\end{enumerate}

\subsection{Science Driven Requirements}

The science and instrument teams distilled the individual science cases discussed above into a set of driving science requirements which GMagAO-X is designed to meet. The resulting key instrument parameters are summarized in Table \ref{tab:params}. The top-level requirements are:

\begin{itemize}[label={}]
    \item \textbf{GMAGX-SCI-001: Exoplanet Characterization in Reflected Light}
    \vspace{-4pt}
    \begin{itemize}[label={}]
        \item \textbf{Requirement: }{GMagAO-X shall measure reflected light from planets around nearby stars such as Proxima Cen and GJ 876.}
    \end{itemize}
\end{itemize}

\begin{itemize}[label={}]
    \item \textbf{GMAGX-SCI-002: Spatial Resolution}
    \vspace{-4pt}
    \begin{itemize}[label={}]
        \item \textbf{Requirement: }{GMagAO-X shall achieve $\lambda/D$ spatial resolution at wavelengths equal and longer than 600 nm on unresolved guide stars.}
        \item \textbf{Goal: }  GMagAO-X shall achieve $\lambda/D$ spatial resolution on resolved guide stars, such as asteroids and resolved stars.
        \item \textbf{Goal: }  GMagAO-X shall achieve $\lambda/D$ spatial resolution at wavelengths longer than 450 nm.
        \item \textbf{Stretch-Goal: }  GMagAO-X shall achieve $\lambda/D$ spatial resolution at wavelengths longer than 350 nm.
    \end{itemize}
\end{itemize}

\begin{itemize}[label={}]
    \item \textbf{GMAGX-SCI-003: Broadband Photometry}
    \vspace{-4pt}
    \begin{itemize}[label={}]
        \item \textbf{Requirement: }{GMagAO-X shall be capable of performing photometry in narrow-band and 10\% bandwidth filter bands.}
    \end{itemize}
\end{itemize}

\begin{itemize}[label={}]
    \item \textbf{GMAGX-SCI-004: Low Resolution Spectroscopy}
    \vspace{-4pt}
    \begin{itemize}[label={}]
        \item \textbf{Requirement: }{GMagAO-X shall be capable of spectroscopy at a resolution between 20-100.}
    \end{itemize}
\end{itemize}

\begin{itemize}[label={}]
    \item \textbf{GMAGX-SCI-005: High Resolution Spectroscopy}
    \vspace{-4pt}
    \begin{itemize}[label={}]
        \item  \textbf{Requirement: }{GMagAO-X shall be capable of spectroscopy at resolutions from 1000-65000.}
        \item \textbf{Goal: } GMagAO-X shall be capable of spectroscopy at resolutions greater than $\mathcal{R}=100,000$ at least in the Oxygen-A band.
    \end{itemize}
\end{itemize}

\begin{itemize}[label={}]
    \item \textbf{GMAGX-SCI-006: Wavelength Coverage}
    \vspace{-4pt}
    \begin{itemize}[label={}]
        \item \textbf{Requirement: }{GMagAO-X shall provide science wavelength coverage from 600-1900 nm to provide discrimination between various exoplanet models.}
        \item \textbf{Goal: }  Wavelength coverage of 450-1900 nm.
        \item \textbf{Stretch Goal: } Wavelength coverage of 350-1900 nm.
    \end{itemize}
\end{itemize}

\begin{itemize}[label={}]
    \item \textbf{GMAGX-SCI-007: Magnitude Range}
    \vspace{-4pt}
    \begin{itemize}[label={}]
        \item \textbf{Requirement: } GMagAO-X shall be capable of observing stars in the magnitude range I = -1.5 to 13.
        \item \textbf{Goal: }  GMagAO-X shall be capable of observing stars in the magnitude range I = -1.5 to 15.
    \end{itemize}
\end{itemize}

\begin{itemize}[label={}]
    \item \textbf{GMAGX-SCI-008: Coronagraph Contrast}
    \vspace{-4pt}
    \begin{itemize}[label={}]
        \item  \textbf{Requirement: } GMagAO-X shall be capable of performing photometry at a signal-to-noise ratio of 5 on a point source with a flux ratio of 1e-7 or better with
                                     respect to its host star in a 10\% bandwidth filter at 4 $\lambda/D$.
        \item \textbf{Goal: }  GMagAO-X shall be capable of performing SNR=5 photometry on a point source with a flux ratio of 1e-7 or better in a 10\% bandwidth filter
                               at 2 $\lambda/D$ and 1e-8 at 6 $\lambda/D$.
        \item \textbf{Stretch Goal: } 1e-8 contrast at 1 $\lambda/D$ and 1e-9 at 5 $\lambda/D$.
    \end{itemize}
\end{itemize}

\begin{itemize}[label={}]
    \item \textbf{GMAGX-SCI-009: Planet Position Measurement}
    \vspace{-4pt}
    \begin{itemize}[label={}]
        \item \textbf{Requirement: }{GMagAO-X shall be capable of measuring companion position with respect to the star with sufficient precision (TBD) to constrain inclination
and the orbital elements.}
    \end{itemize}
\end{itemize}

\begin{itemize}[label={}]
    \item \textbf{GMAGX-SCI-010: Field of View}
    \vspace{-4pt}
    \begin{itemize}[label={}]
        \item \textbf{Requirement: }{GMagAO-X shall provide a field of view of at least 3 arcseconds x 3 arcseconds.}
    \end{itemize}
\end{itemize}

\begin{itemize}[label={}]
    \item \textbf{GMAGX-SCI-011: Relative Photometric Stability}
    \vspace{-4pt}
    \begin{itemize}[label={}]
        \item  \textbf{Requirement: }{GMagAO-X shall be capable of 1\% (TBC) relative photometric stability over a 4 hr (TBC) period in coronagraphic high-contrast observations.}
    \end{itemize}
\end{itemize}

The \textit{GMagAO-X Instrument Science Requirements Document} (GMAGX-DOC-0001) contains much more discussion about the rationale for each of these 11 requirements and how they
flow down from the science cases.  These requirements flow down to the instrument level requirements, on which the preliminary design of GMagAO-X is based.

\section{PRELIMINARY DESIGN}

\subsection{MECHANICAL DESIGN}

\begin{figure} [h!]
\centering
\includegraphics[width=4.5in]{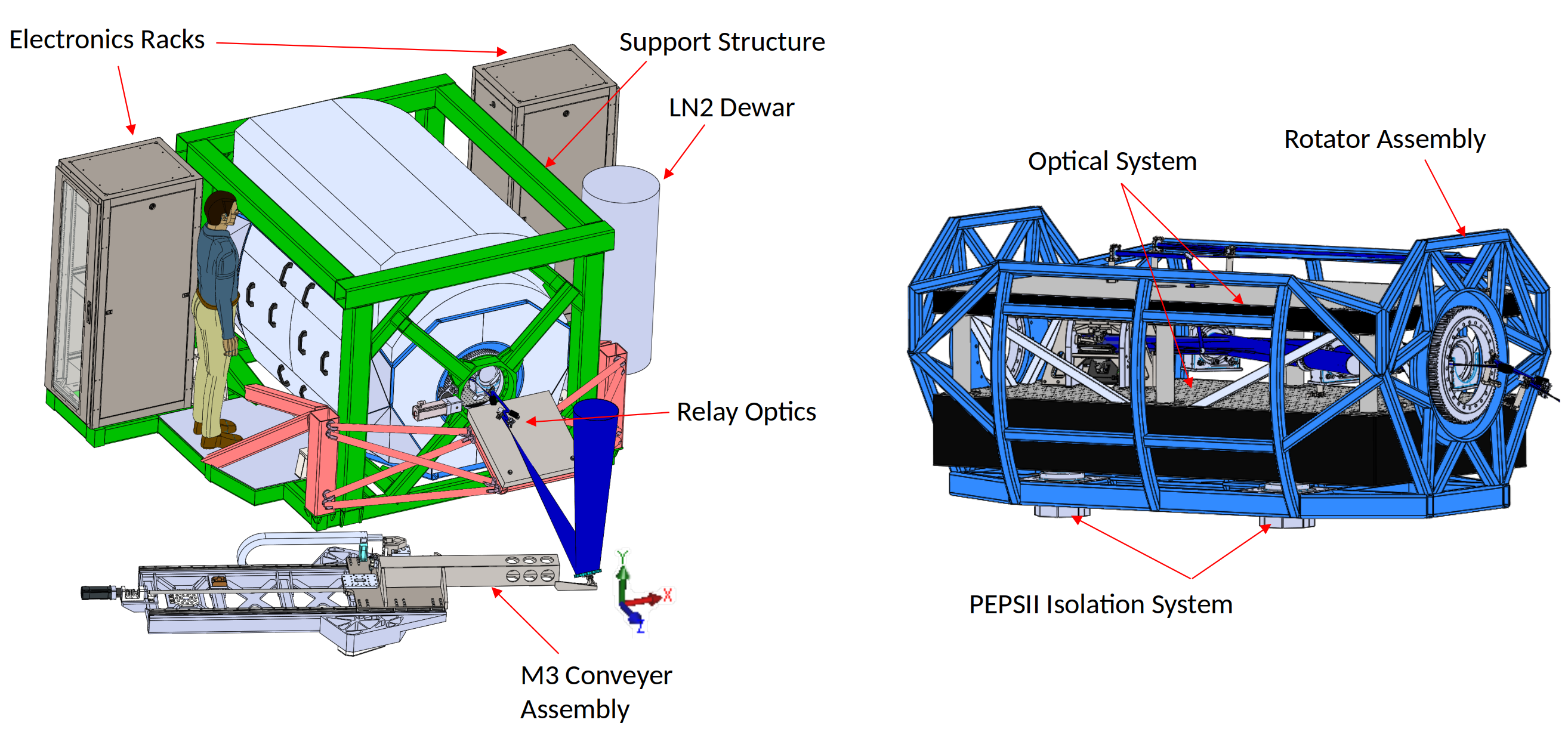}
\caption{\label{fig:mech1} Overview of the Mechanical Design of GMagAO-X}
\end{figure}

The mechanical design of GMagAO-X is shown in Figure \ref{fig:mech1}.  GMagAO-X will occupy a Folded Port (FP) on the Gregorian Instrument Rotator (GIR) of the GMT. The main structure of GMagAO-X consists of a
small optical bench with fore-optics and a fast-steering mirror, a rotating frame containing the main two-level optical subsystem, and onboard electronics racks which primarily house DM electronics and so must be close
to the instrument.  GMagAO-X also supplies its own custom tertiary mirror (M3).  This decision was made to enable a very high quality, but smaller than facility-sized, optic to be used for M3.

To provide vibration isolation, the two-level optical table is floating (air-damped).  We will use an actively controlled, in height and level, system from TMC called ``PEPSII'', which we have demonstrated with the existing MagAO-X instrument on the Magellan Clay telescope.  See Close et al.\cite{close_2024} in these proceedings.

\begin{figure} [h!]
\centering
\includegraphics[width=4.5in]{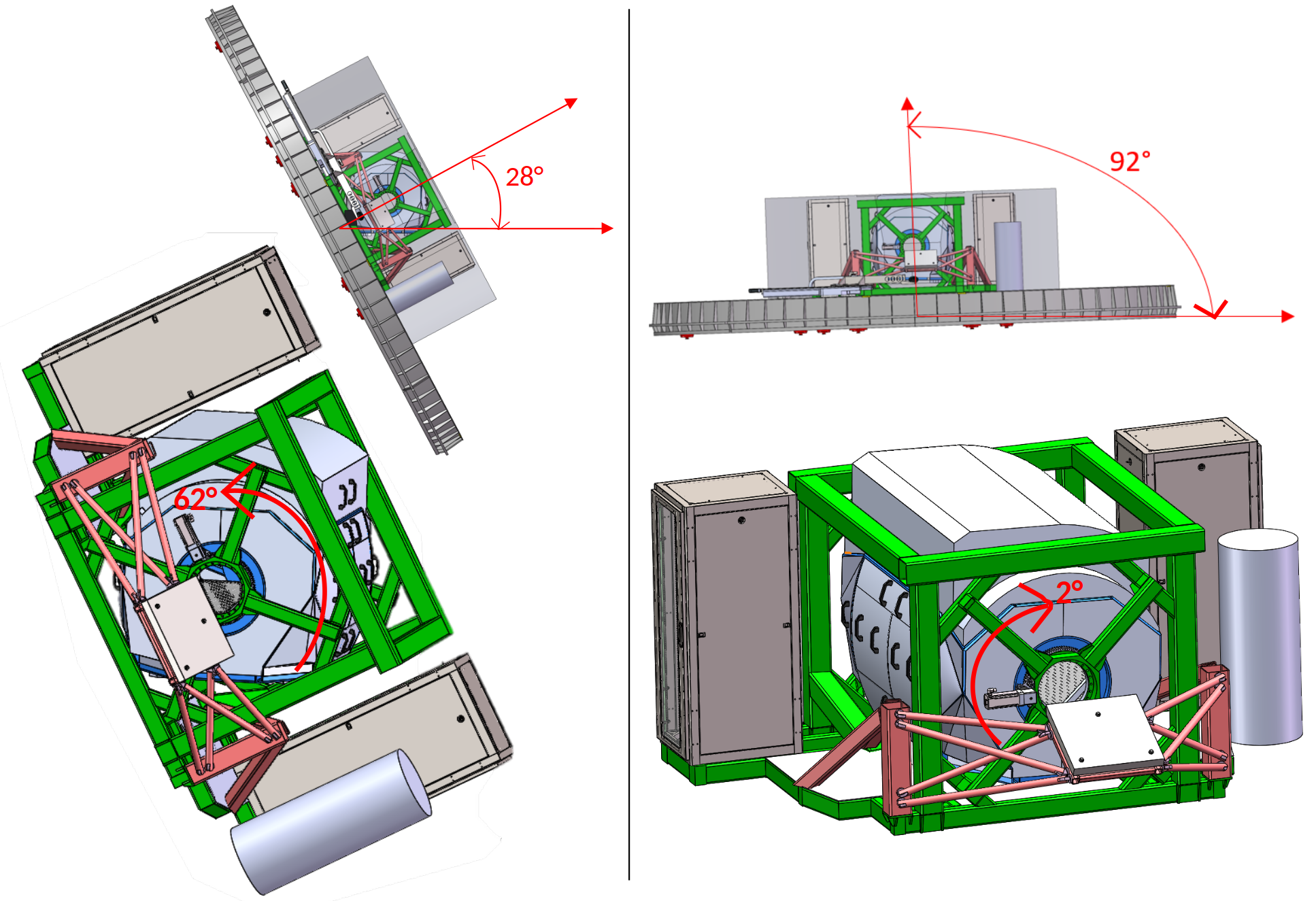}
\caption{\label{fig:mech2} Gravity Invariance: The optical table of GMagAO-X is rotated to be gravity invariant during operation. }
\end{figure}

\begin{figure} [h!]
\centering
\includegraphics[width=4.5in]{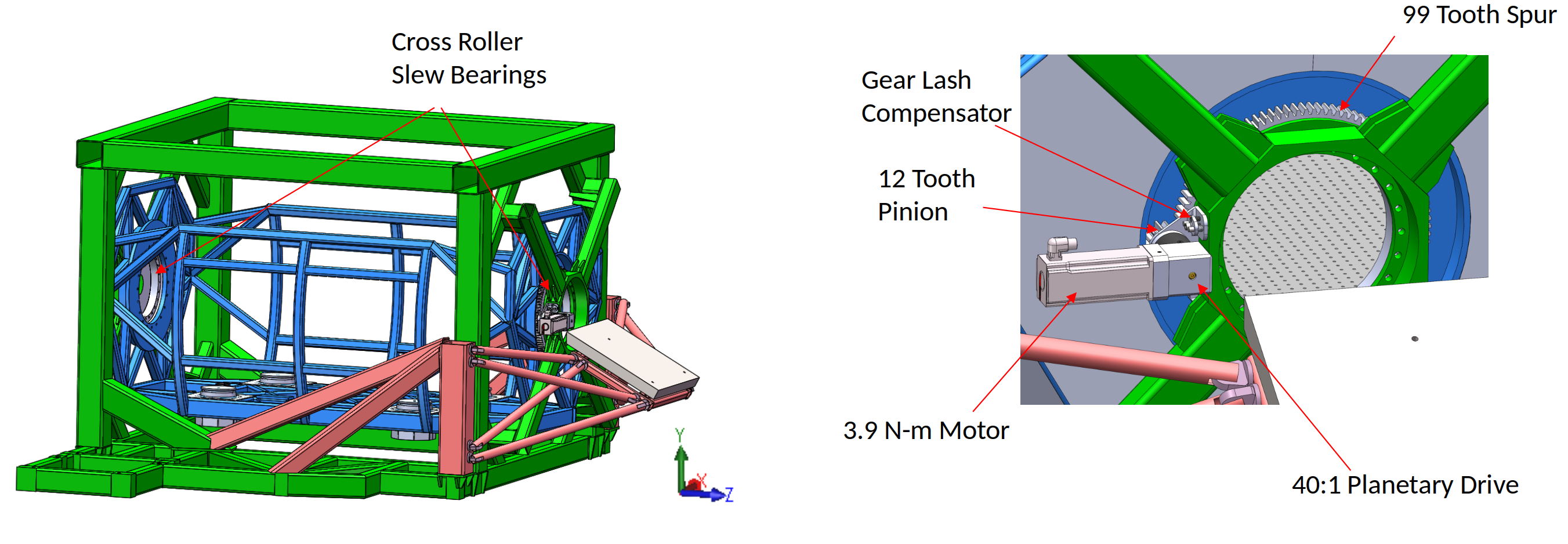}
\caption{\label{fig:mech3} The Rotation Mechanism of GMagAO-X}
\end{figure}

The GIR is below the primary mirror (M1), and tips with elevation.  The GIR, as its names implies, rotates to track the sky.  To maximize stability and provide the ability to use pneumatic vibration isolation, GMagAO-X is designed to counter rotate such that its main optics are gravity invariant (see Figure \ref{fig:mech2}).  To facilitate this, when GMagAO-X is in operation the GIR will be locked in position such that GMagAO-X will be parallel to the elevation axis of the GMT.  Figure \ref{fig:mech3} illustrates the crossed roller bearing and drive system for rotating the optical table.

When not in operation, GMagAO-X will not be floating and is designed to travel through the full range of motion of the telescope.  This includes a locking mechanism to clamp the table down rigidly and ensure that it is constrained for all possible angles of the GIR while other instruments are in operation.

The preliminary mechanical design of GMagAO-X has been thoroughly analyzed for structural integrity, and to ensure it meets the demanding seismic survivability standards of the GMT.  All aspects of the construction, shipment, assembly, installation and removal, alignment, and night-time operations have been considered in the design.

\subsection{OPTICAL DESIGN}

\begin{figure} [b]
\centering
\includegraphics[width=6.5in]{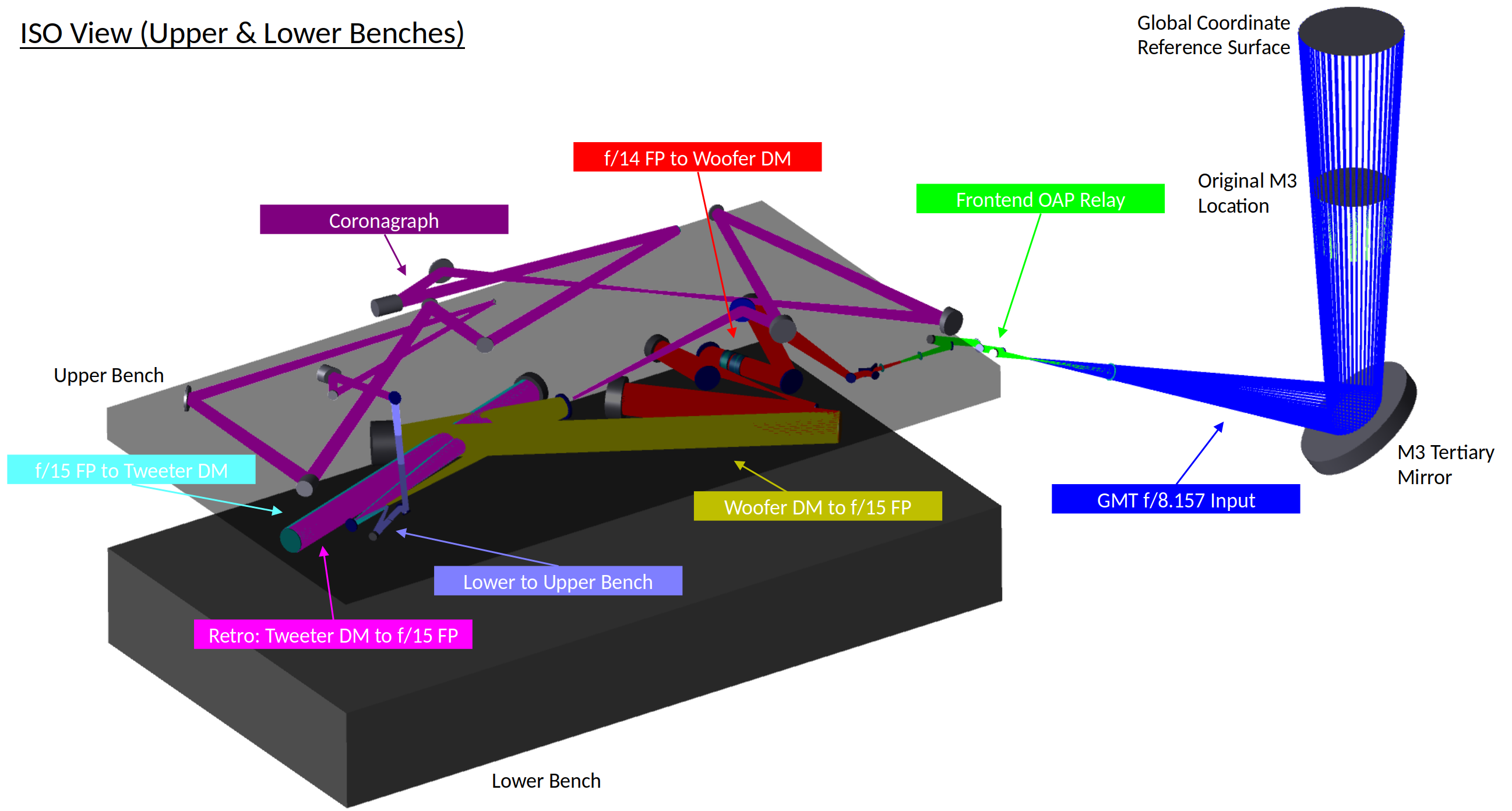}
\caption{\label{fig:opt1} The Optical Design of GMagAO-X}
\end{figure}

Figure \ref{fig:opt1} shows a summary of the optical design of GMagAO-X.  The custom M3 reflects the f/8.157 beam of the GMT into the frontend OAP Relay (green).  These fore-optics contain a fast steering mirror (FSM) to provide high-stroke and high-speed vibration control.  The beam next passes through a pupil alignment periscope, a k-mirror de-rotator, and atmospheric dispersion corrector (ADC).  A 3000 actuator DM ``woofer'' is used to provide large-stroke for correcting low-order aberrations at lower speeds.  After the woofer, the beam is relayed to the 21,000 ``parallel DM'' which provides the high-speed and high-spatial-frequency control.  The beam is then split between the wavefront sensors (WFS) and the coronagraph.  The wavefront sensing subsystem (not shown in Figure \ref{fig:opt1} contains both a Holographic Dispersed Fringe Sensor (HDFS,\cite{10.1117/1.JATIS.8.2.021513,10.1117/1.JATIS.8.2.021515}) for coarse phasing control and a high-order WFS.  A 3-sided pyramid WFS is currently baselined (see Haffert et al.\cite{haffert_2024b} in these proceedings).

The coronagraph beam is relayed to the upper level (purple rays in Figure \ref{fig:opt1}.  The coronagraph contains a 3000 actuator non-common path correcting (NCPC) DM, an architecture which has been demonstrated on-sky with MagAO-X\cite{2022SPIE12185E..09M,males_2024}.  This allows in-coronagraph WFS\&C, such as the digging of dark holes\cite{2023A&A...673A..28H,haffert_2024}, without offsetting to the high-order WFS\&C system.  After the NCPC DM, the coronagraph optics provide a Lyot-style architecture.  The baseline coronagraph is the Phase-Apodized Pupil Lyot Knife-Edge Coronagraph (PAPLKEC)\cite{2020ApJ...888..127P}, which uses the NCPC DM for apodization and a knife-edge mirror as the focal plane mask.  To push towards the goal and stretch-goal of GMAGX-SCI-008  the design include Phase Induced Amplitude Apodization (PIAA\cite{2003A&A...404..379G}) optics (including inverse PIAA) and transmissive complex focal plane masks.  Both focal-plane low-order WFS (FLOWFS) and Lyot-plane LOWFS (LLOWFS) are supported to make use of light rejected by the coronagraphs.

After the coronagraph, focal plane instrumentation will include imagers covering the optical through 1 $\mu$m, and near-IR through H band (see GMAGX-SCI-006).   We plan to couple GMagAO-X to the facility G-CLEF spectrograph\cite{szentgyorgyi_2024} as a fiber fed integral field unit (IFU), and possibly to other facility spectrographs for near-IR coverage.  We have also left room for an on-board IFU which is yet to be designed.

See Close et al.\cite{close_2024} in these proceedings for further details about the optical design of GMagAO-X.

\subsubsection{The Parallel DM}

\begin{figure} [h!]
\centering
\includegraphics[width=5in]{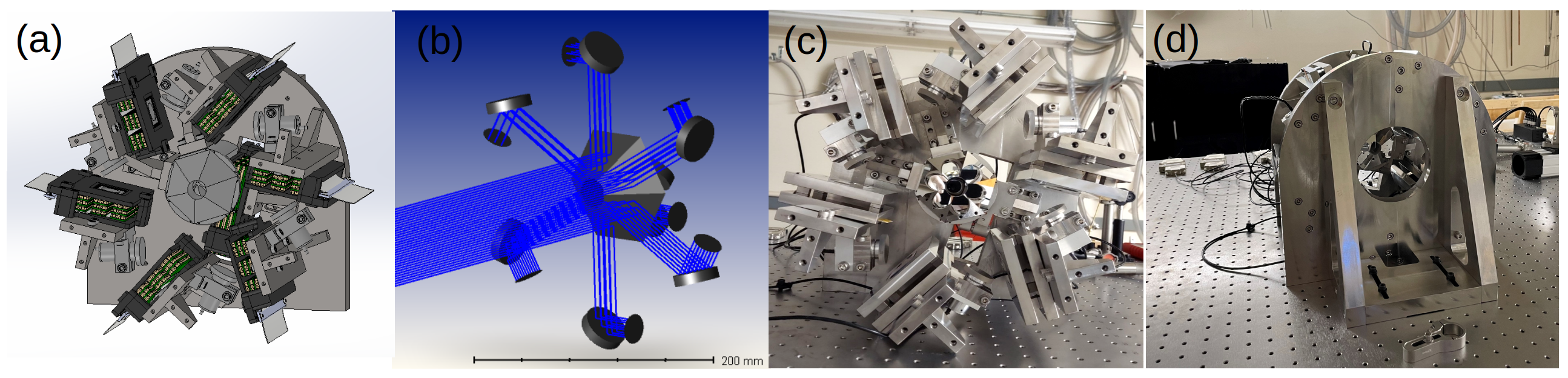}
\caption{\label{fig:opt2} The Parallel DM Concept: the large-segment primary mirror of GMT motivates using a segmented DM with 3000 actuators per segment. (a) Mechanical Design. (b) Optical Design.  Note the crossed folds for polarization control. (c) \& (d) Prototype assembled and tested on the HCAT testbed at the University of Arizona.}
\end{figure}

The key to achieving the WFS\&C required for the science goals of GMagAO-X is the ability to deploy a high-actuator count DM.  Based on the specifications and performance of MagAO-X\cite{2022SPIE12185E..09M,males_2024}, a $\sim$14 cm projected pitch is needed, which is equivalent to $\sim$3000 illuminated actuators per segment.  Figure \ref{fig:opt2} shows how this can be achieved on the GMT with the ``Parallel DM'' architecture.  We use a hexagonal prism with its 6 segments coated as mirrors.  Placed near a pupil plane, this splits the GMT aperture such that each segment is imaged onto a 3K MEMS DM.  The central hole passes the center segment to its DM in the back.  Crossed folds are used to control polarization effects, and folds are actuated with piston-tip-tilt stages to provide coarse phasing control.  Fine high-speed phasing control is provided by the MEMS segments.  Figure \ref{fig:opt2} shows how the parallel DM fits on the lower table of GMagAO-X.

The parallel DM has been prototyped and demonstrated, albeit without MEMS installed, in the High Contrast Adaptive Optics phasing Testbed (HCAT) at the University of Arizona.  See Close et al.\cite{close_2024} in these proceedings and Kautz et al (submitted to JATIS).

\begin{figure} [h!]
\centering
\includegraphics[width=3in]{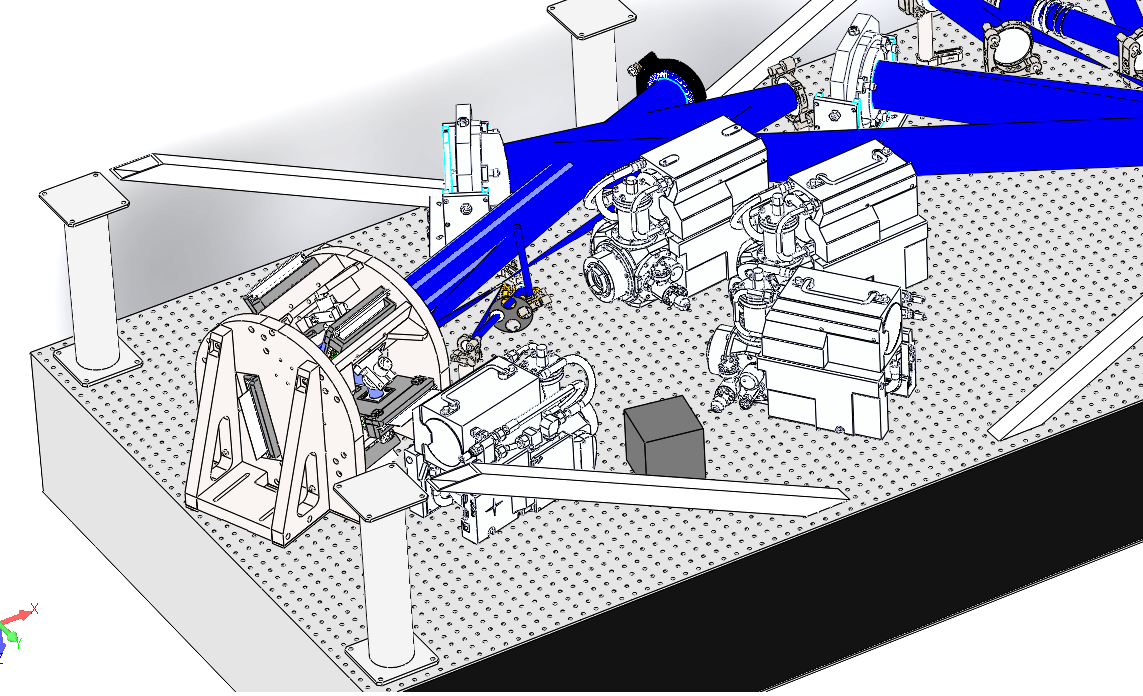}
\caption{\label{fig:opt3} The parallel DM is shown rendered in the opto-mechanical design of GMagAO-X.}
\end{figure}

\subsubsection{Coronagraphs}

\begin{figure}[h]
\centering
\includegraphics[width=3in]{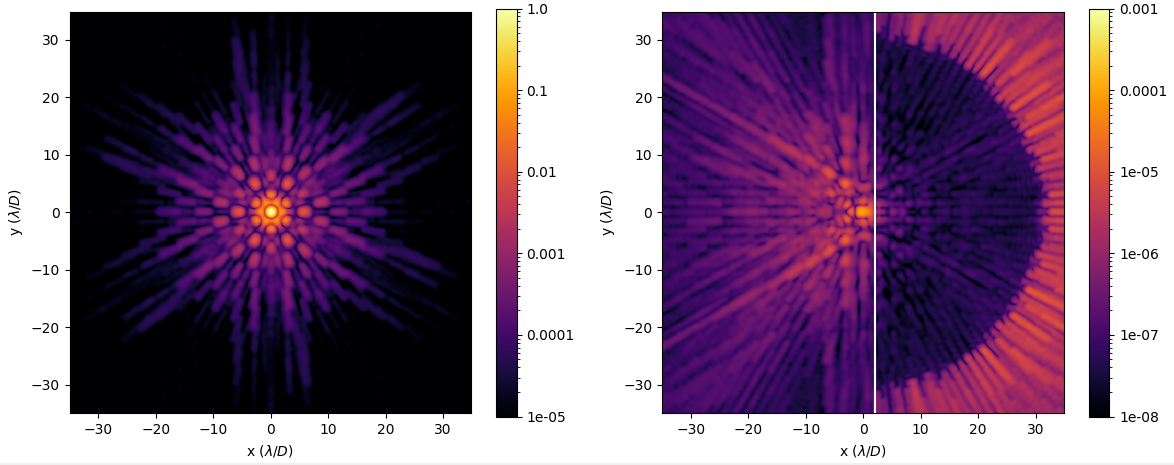}
\includegraphics[width=3in]{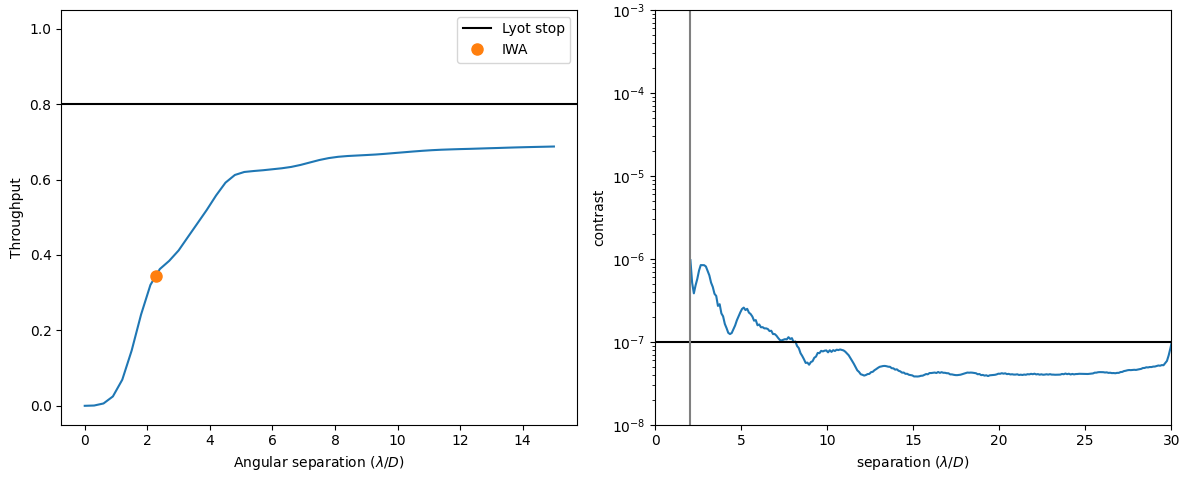}
\caption{Preliminary PAPLKEC coronagraph design for GMagAO-X.  This design is for a 20\% bandwidth.  Left shows the PSF. Middle-left shows the dark hole (with no residual atmosphere).
Middle-right shows the throughput curve, with inner working angle of 2.3 $\lambda/D$ defined at 50\% throughput.  Right shows raw contrast vs. separation.  \label{fig:corons}}
\end{figure}

Figure \ref{fig:corons} documents the baseline PAPLKEC coronagraph design.  The design easily supports achieving the science goals define above.  The GMT aperture presents no significant challenges for
modern coronagraph designs.

\subsection{ELECTRICAL DESIGN}

\begin{figure} [h!]
\centering
\includegraphics[width=6.5in]{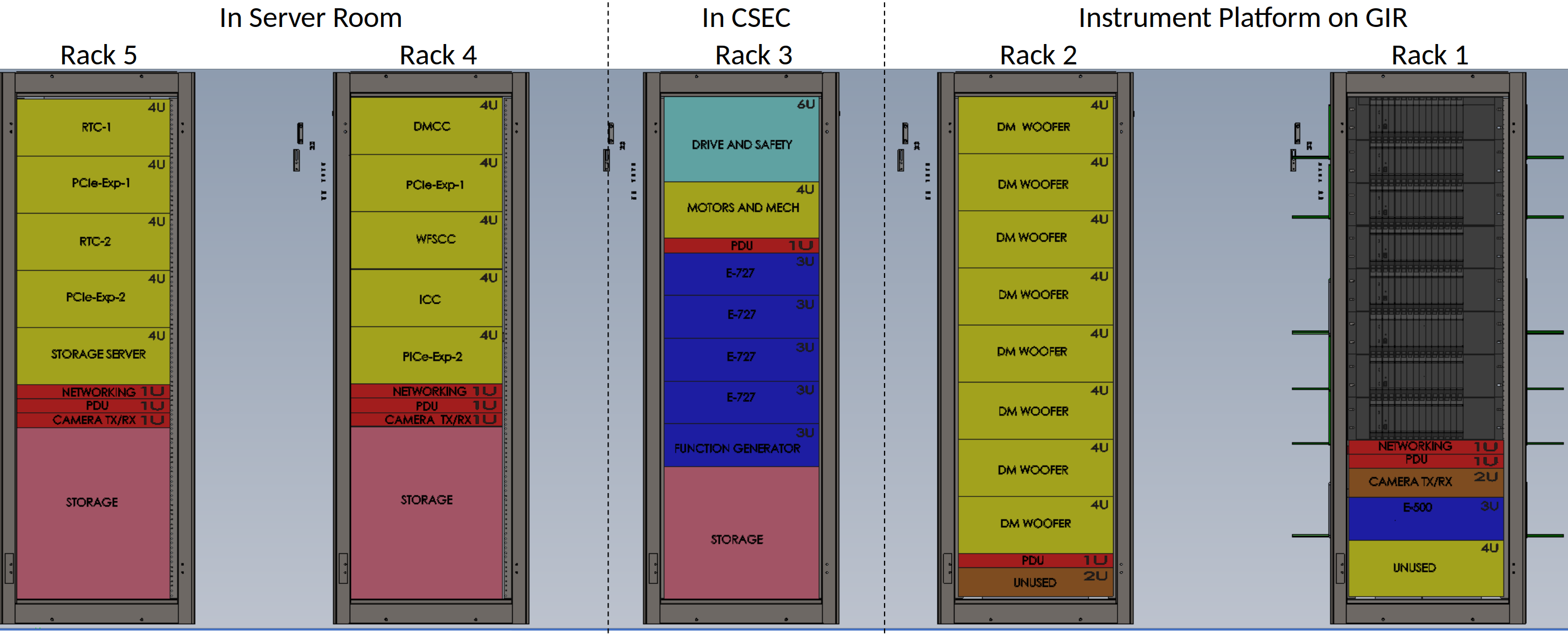}
\caption{\label{fig:elec1} The equipment racks used to house GMagAO-X electronics. Rack 1 houses the 8 MEMS driver (7x for the parallel DM tweeter, 1x for the NCPC DM and image acquisition hardware.  Rack 2 houses the 32 U ALPAO DM 3228 driver.  Rack 3 houses motion control and drivers for the parallel DM PTT stages.  Racks 4 and 5 are off the telescope and house the distributed real-time control computer system.}
\end{figure}

The electronics of GMagAO-X are concerned with the safe operation of the rotation system and the air floating system, as well as the operation of the ExAO system and coronagraph components.    The main feature
of the electrical design is the layout of the equipment racks that will house the drive and control electronics.  Figure \ref{fig:elec1} illustrates the 5 racks we plan to use.

Two racks will be mounted on the instrument frame.  These are primarily occupied by DM driver electronics, which due to cable length restrictions must be as close as possible to the devices.  Other components with similar cable length and/or latency restrictions are included in these racks.  Note that these racks tip with elevation, but do not rotate with the internal barrel and so all cabling from
these racks is managed by a cable guide.  A third rack below the deck of the GIR, co-rotating with azimuth, will house additional motion control electronics and networking gear.  Power distribution is distributed
amongst the racks.  The entire power requirement of GMagAO-X is well within the allotment for an FP instrument, as is the thermal performance and cooling requirements.

At least two additional racks will be located off the telescope in an equipment/computer room, to house the distributed real-time control computer system.  This is described further below.  Sufficient allocations
of optical fiber have been reserved between the GIR and the equipment room to support GMagAO-X's unique needs.

\begin{figure} [h!]
\centering
\includegraphics[width=4.5in]{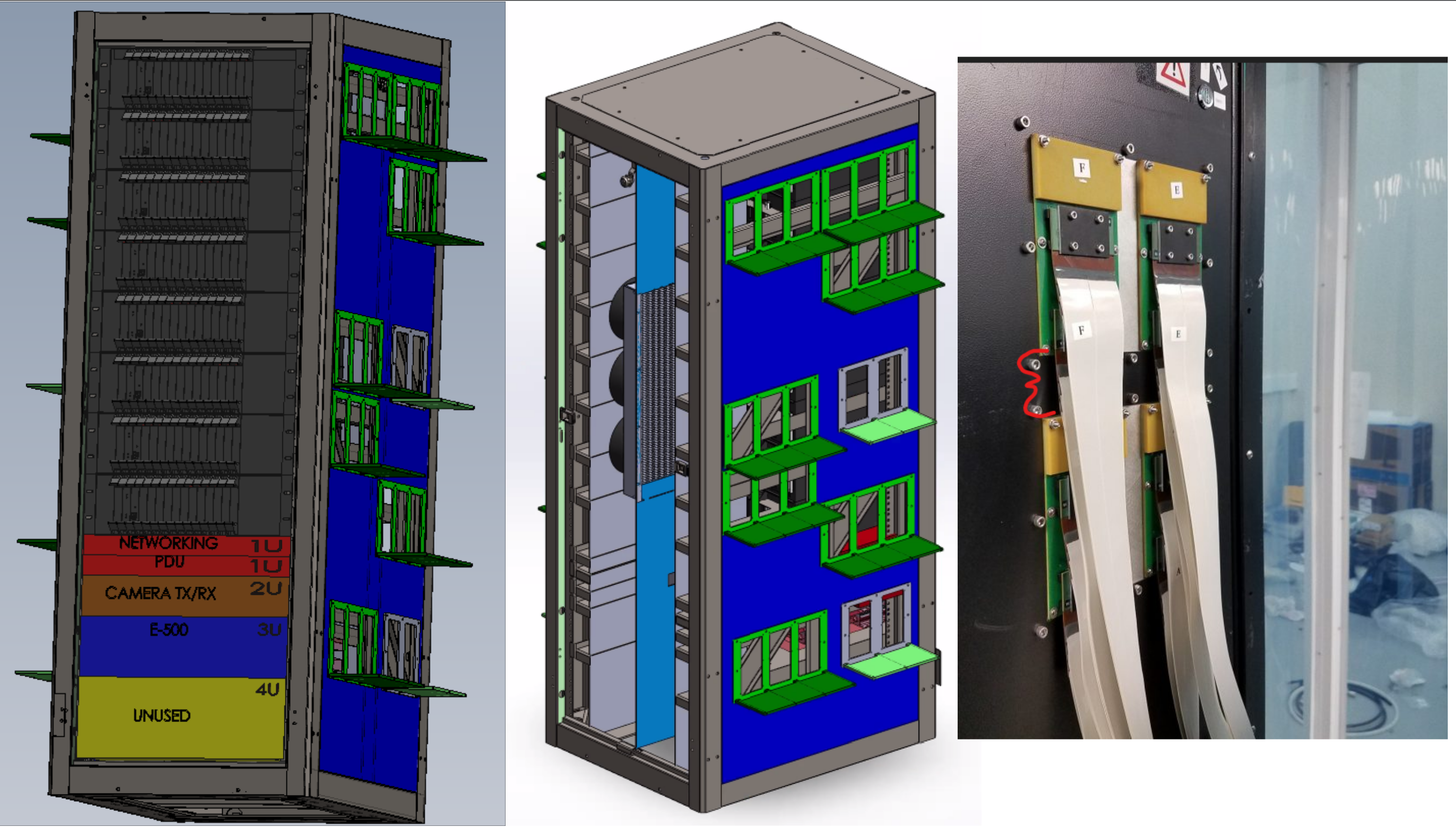}
\caption{\label{fig:elec2} Management of MEMS ribbon cables.  Feed through connectors will be placed on the outside rack \#1 housing the MEMS electronics.  The right photo shows such a system
in operation on the current MagAO-X instrument.}
\end{figure}

An important concern with MEMS DM is cable management, as the ribbon cables typically used to supply per-actuator voltages take up significant space.  In order to support occasional maintenance and shipping,
we will provide bulkhead connectors on the outside of the MEMS rack, as shown in Figure \ref{fig:elec2}.  This will allow the rack to be disconnected from the instrument without disturbing either the
electronics or the DM segments.  We have extensive experience with this system on MagAO-X, which undergoes routine shipping and transportation on the mountain.

\subsection{SOFTWARE \& CONTROL SYSTEM}

The baseline software system for GMagAO-X is based on the proven MagAO-X software suite.  The MagAO-X software is based on a modern C++ API, with a base class inherited by all applications which
implements standard housekeeping (startup sequence, event loop, logs, telemetry, and interprocess communication (IPC, both routine and low-latency)).  We use a binary logging framework for efficiency.
The architecture is multi-application, with multiple threads in each application.  The Instrument Neutral Distributed Interface (INDI) protocol is used for soft-real-time IPC, and the ImageStreamIO low-latency
library is used for WFS\&C image processing and hard-real-time IPC. The MagAO-X framework is robust and fully demonstrated on-sky, as well as being used in several lab testbeds, and in addition,  significant effort is underway to prepare
it for space-flight applications.  The MagAO-X PDR and pre-ship review (PSR) software designs are available online\footnote{\url{https://magao-x.org/docs/handbook/_downloads/78a8f3b30b90bdcb2a4560f4c0981fca/3.3_software.pdf}}
\footnote{\url{https://magao-x.org/docs/handbook/_downloads/e9e6b896c53ab5b359d890763de06e96/2_5_Software_Processes.pdf}}.  The API reference\footnote{\url{https://magao-x.org/docs/api/}}
and source code are also public\footnote{\url{https://github.com/magao-x/MagAOX}}.

The real-time control system for GMagAO-X is based on the Compute and Control for Adaptive Optics (CACAO\cite{2018SPIE10703E..1EG}) system. CACAO is in routine use at SCExAO\cite{2015PASP..127..890J} and on MagAO-X.
The shared memory architecture of CACAO (MILK ImageStreamIO) is the backbone of our image transfer and low-latency IPC.
We use this for the science cameras as well, enabling low-latency focal plane WFS (FPWFS).

\begin{figure} [h!]
\centering
\includegraphics[width=3.25in]{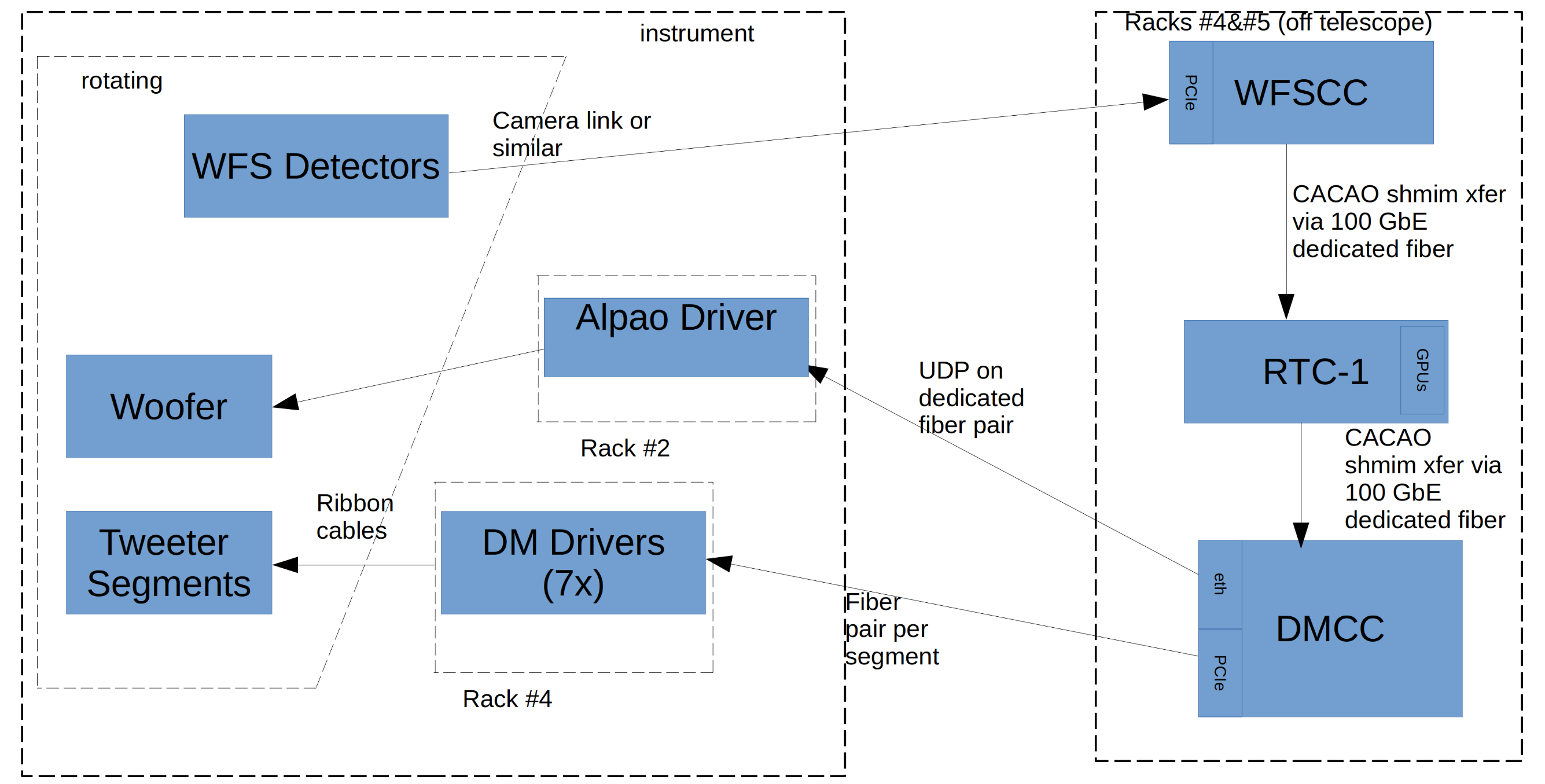}
\includegraphics[width=3.25in]{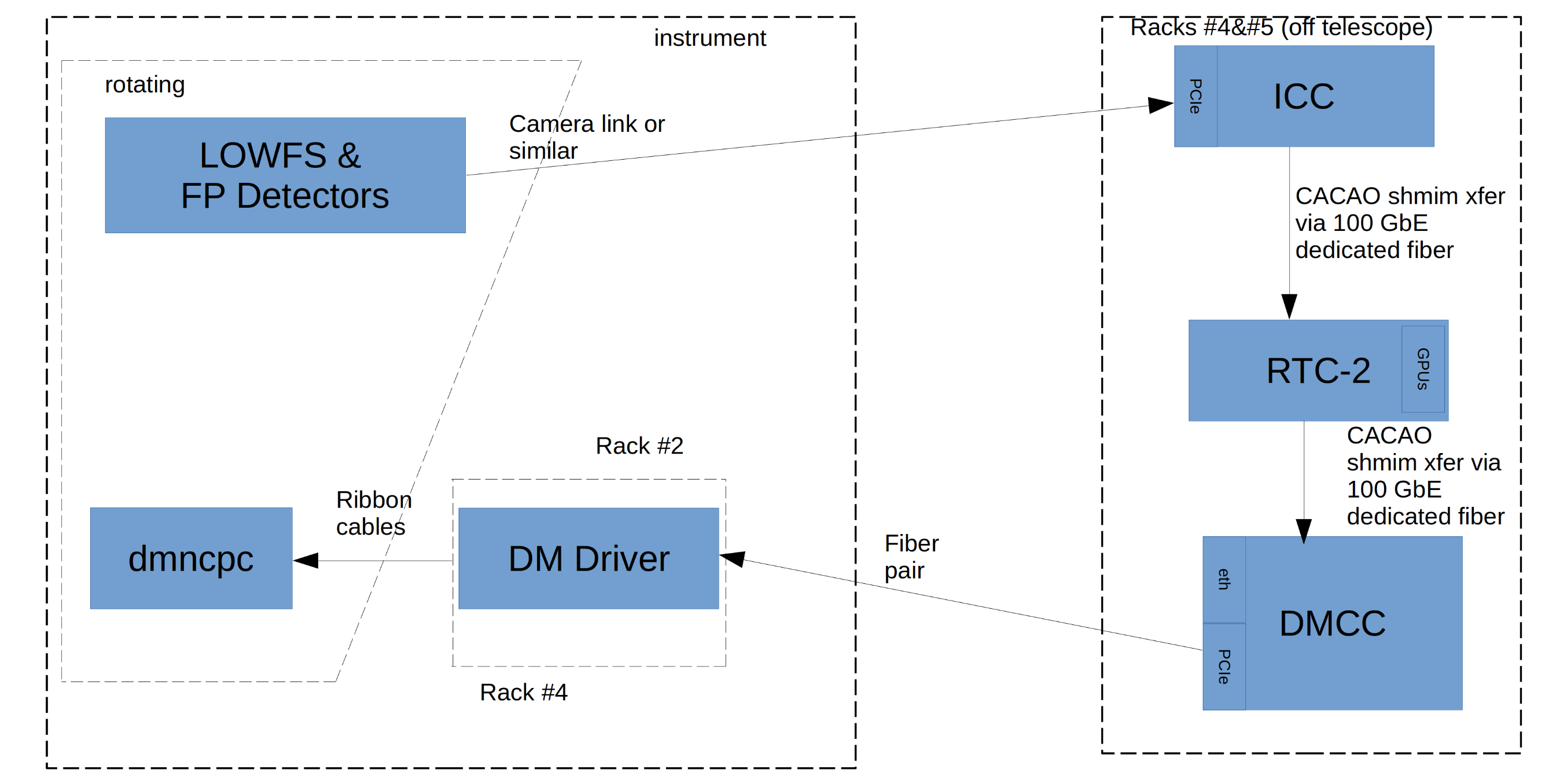}
\caption{\label{fig:comp1} The distributed real-time control system of GMagAO-X}
\end{figure}

In order to enable image acquisition from and control of the large number of devices in GMagAO-X, and to enable the processing throughput needed for a 21,000 actuator AO system, we plan to
implement a distributed real-time control system.  A key component of CACAO that is needed for GMagAO-X is low-latency
computer-to-computer image transfer.  This is also used routinely on SCExAO and MagAO-X for wavefront control.  The computers and their task allocation are:
\begin{itemize}
    \item WFSCC: WFS Control Computer
    \vspace{-4pt}
        \begin{itemize}
            \item Low-latency camera readout (one PCIe slot per camera, dedicated signal line(s) each (cameralink, coaxpress))
            \item Initial calibrations (dark sub, flat field, masking) applied here before transfer
            \item Low-latency transfer to RTC-1
            \item Ancillary electronics control (modulator, shutters, filters, etc.) 
        \end{itemize}
    \item ICC: Instrument Control Computer
    \vspace{-4pt}
        \begin{itemize}
            \item Low-latency camera readout  (one PCIe slot per camera, dedicated signal line(s) each (cameralink, coaxpress))
            \item Low-latency transfer to RTC-2
            \item Ancillary electronics control (shutters, filters, etc.) 
        \end{itemize}
    \item RTC-1: Real-Time Computer for Main AO
    \vspace{-4pt}
        \begin{itemize}
            \item Low-latency transfer from WFSCC, to DMCC
            \item PCIe expansion (GPUs)
        \end{itemize}
    \item RTC-2: Real-Time Computer for Coronagraph AO
    \vspace{-4pt}
        \begin{itemize}
            \item Low-latency transfer from ICC, to DMCC
            \item PCIe expansion (GPUs)
        \end{itemize}
    \item DMCC: DM Control Computer
    \vspace{-4pt}
        \begin{itemize}
            \item Commands to DM drivers (one PCIe per segment, dedicated fiber pair each)
            \item Low-latency transfer from RTC-1 \& RTC-2
        \end{itemize}
\end{itemize}
We make use of PCIe expansion to provide the data acquisition, control output, and computing capacity needed.  Figure \ref{fig:comp1} illustrates the layout of the GMagAO-X control system.

\section{PERFORMANCE \& Exoplanet Yield}

\begin{figure}
\centering
\includegraphics[width=2.5in]{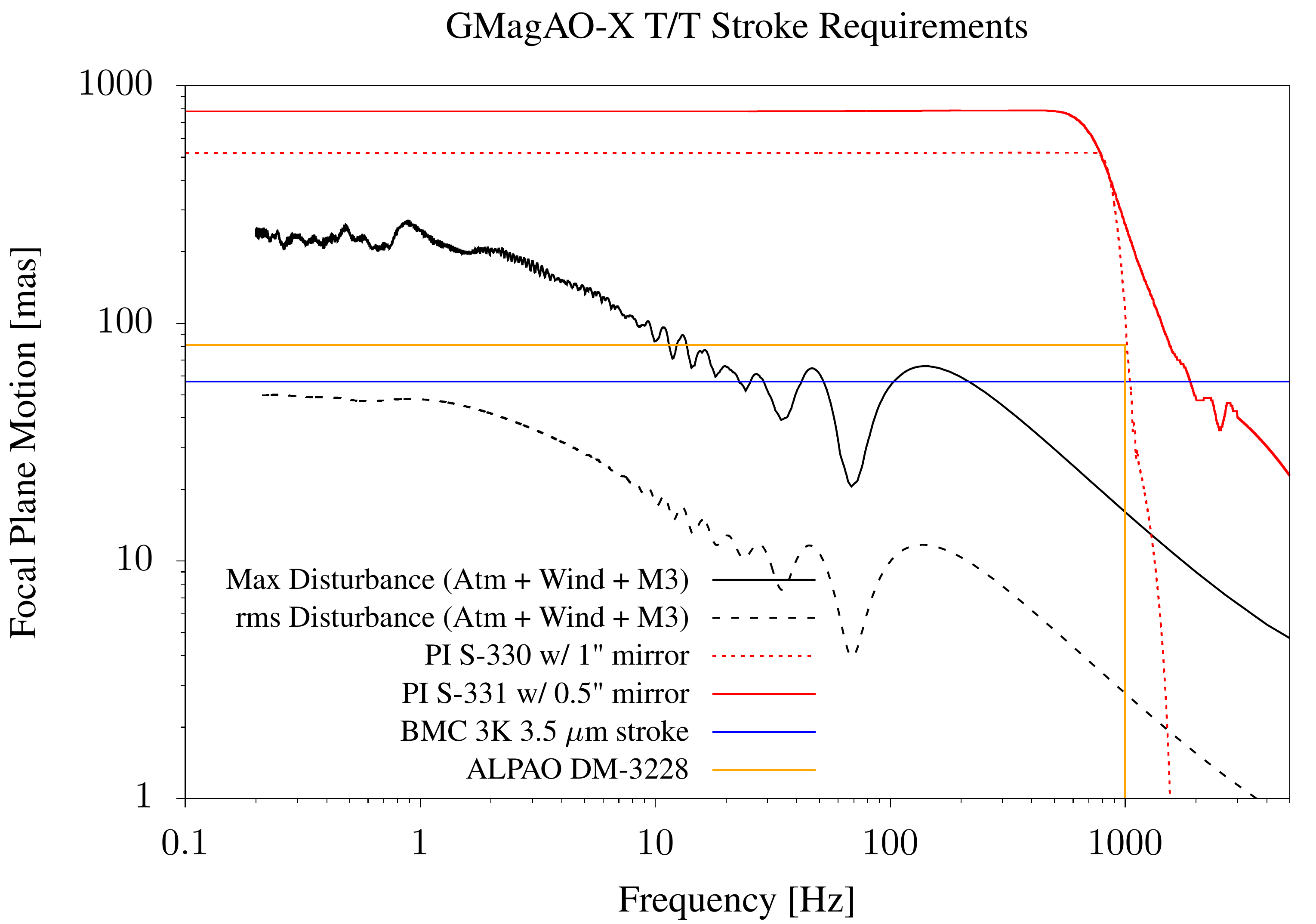}
\includegraphics[width=2.5in]{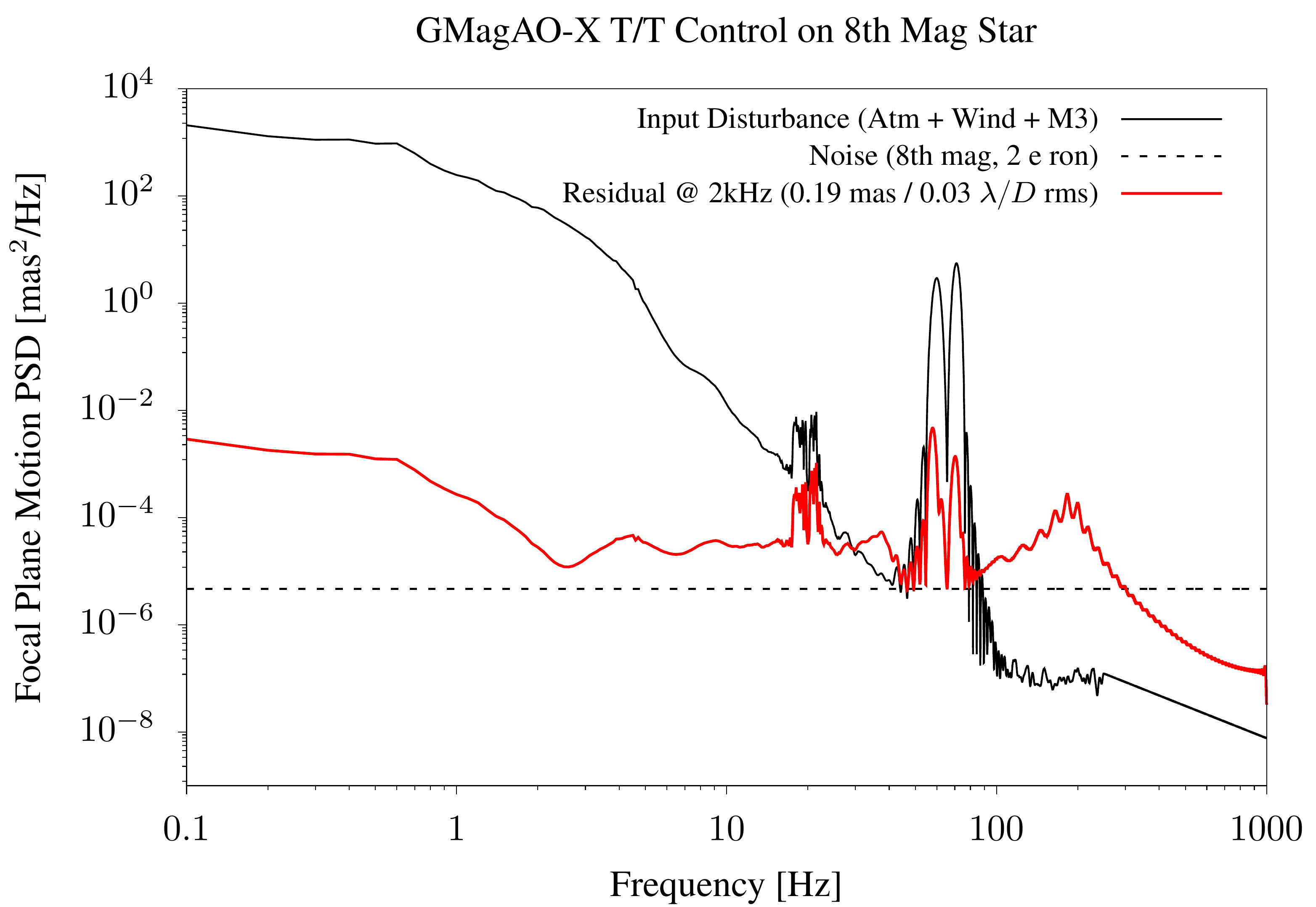}
\caption{Left: GMagAO-X tip/tilt stroke requirements.  Right: GMagAO-X tip/tilt suppression on an 8th mag star \label{fig:tt}}
\end{figure}

We used dynamical models of the GMagAO-X instrument to assess the performance of the preliminary design.  Figure \ref{fig:tt} shows the tip/tilt \& vibration control performance of GMagAO-X.
This includes the atmosphere, the results of the structural modeling done by the GMT, contributions from GMagAO-X itself (dominated by M3), and the results of closed-loop control.
The FSM (a PI S-331 is the baseline) has sufficient stroke to correct the worst case  tip/tilt, and the resulting $0.03 \lambda/D$ rms WFE residual meets the instrument level requirement set by coronagraph performance.

\begin{figure}
\centering
\includegraphics[width=2.5in]{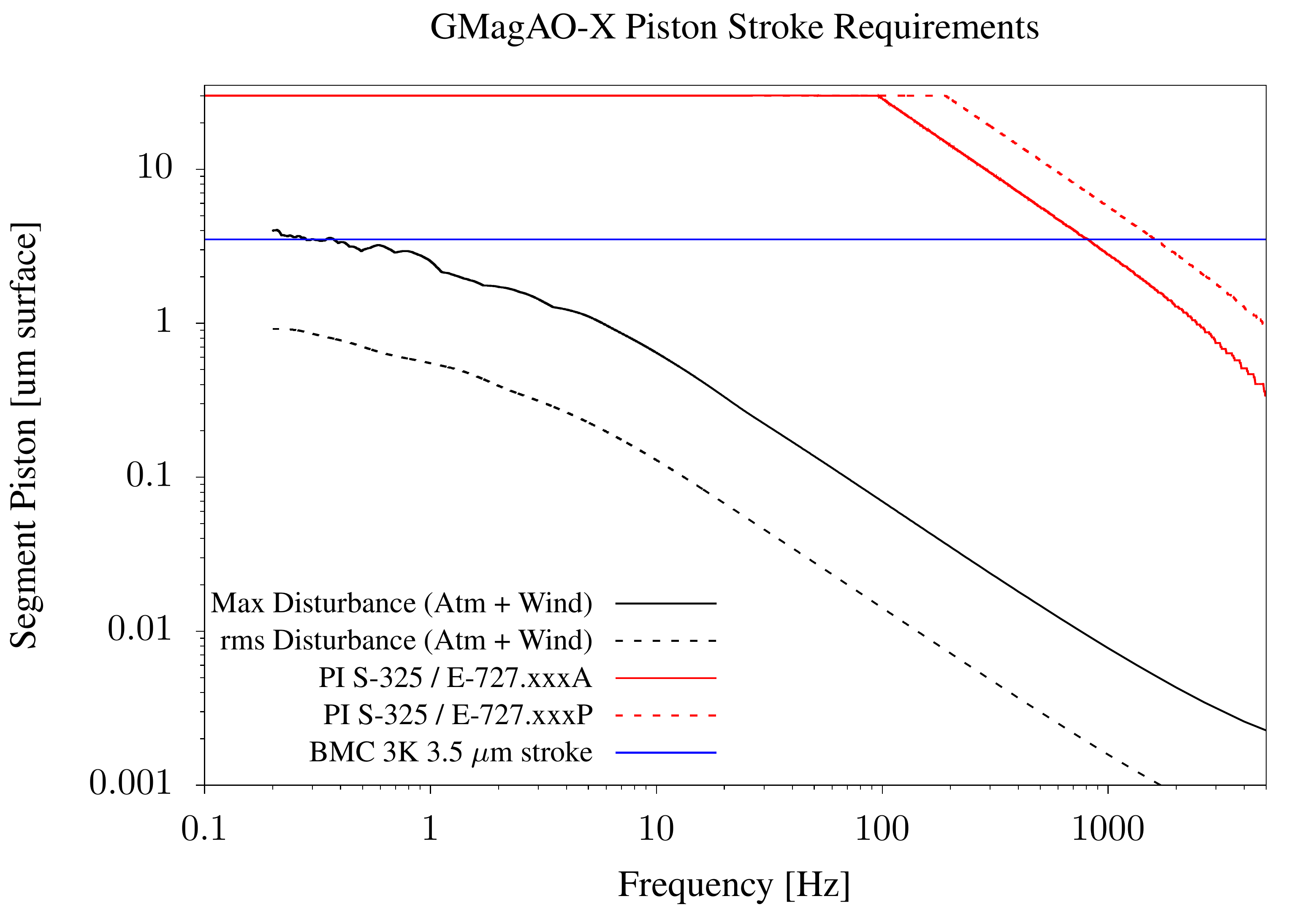}
\includegraphics[width=2.5in]{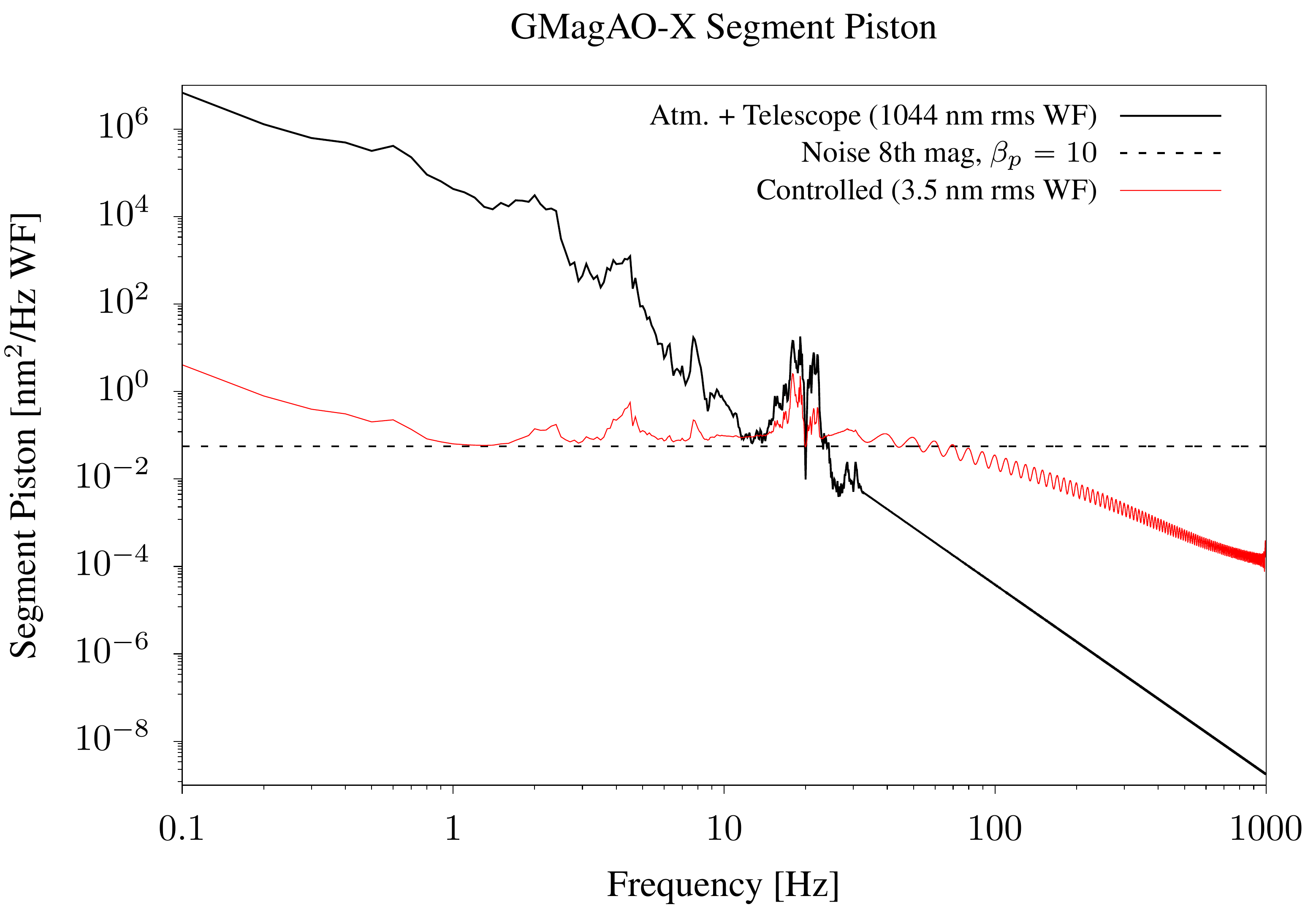}
\caption{Left: GMagAO-X segment piston stroke requirements.  Right: GMagAO-X segment piston suppression on an 8th mag star \label{fig:piston}}
\end{figure}

The same analysis was conducted for segment piston. Figure \ref{fig:piston} shows the segment piston control performance of GMagAO-X.  This includes the atmosphere, the results of the structural
modeling done by the GMT,  and the results of closed-loop control. The resultant 3.5 nm rms WFE residual meets the instrument level requirement set by coronagraph performance.

\begin{figure}
\centering
\includegraphics[width=2.5in]{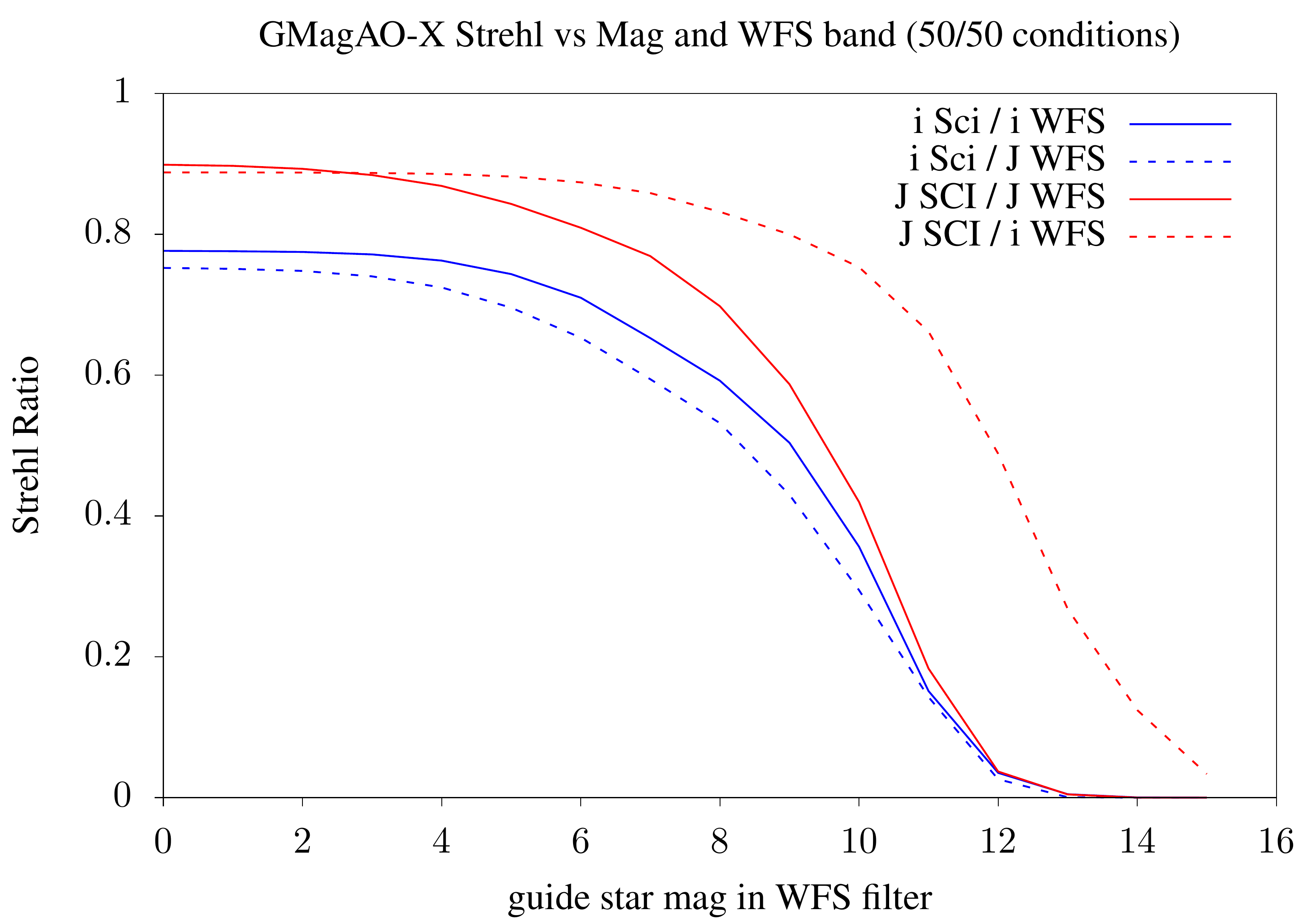}
\caption{Strehl ratio vs. guide star magnitude for GMagAO-X, for different combinations of WFS and science wavelengths. \label{fig:strehl}}
\end{figure}

\begin{figure}
\centering
\includegraphics[width=2.5in]{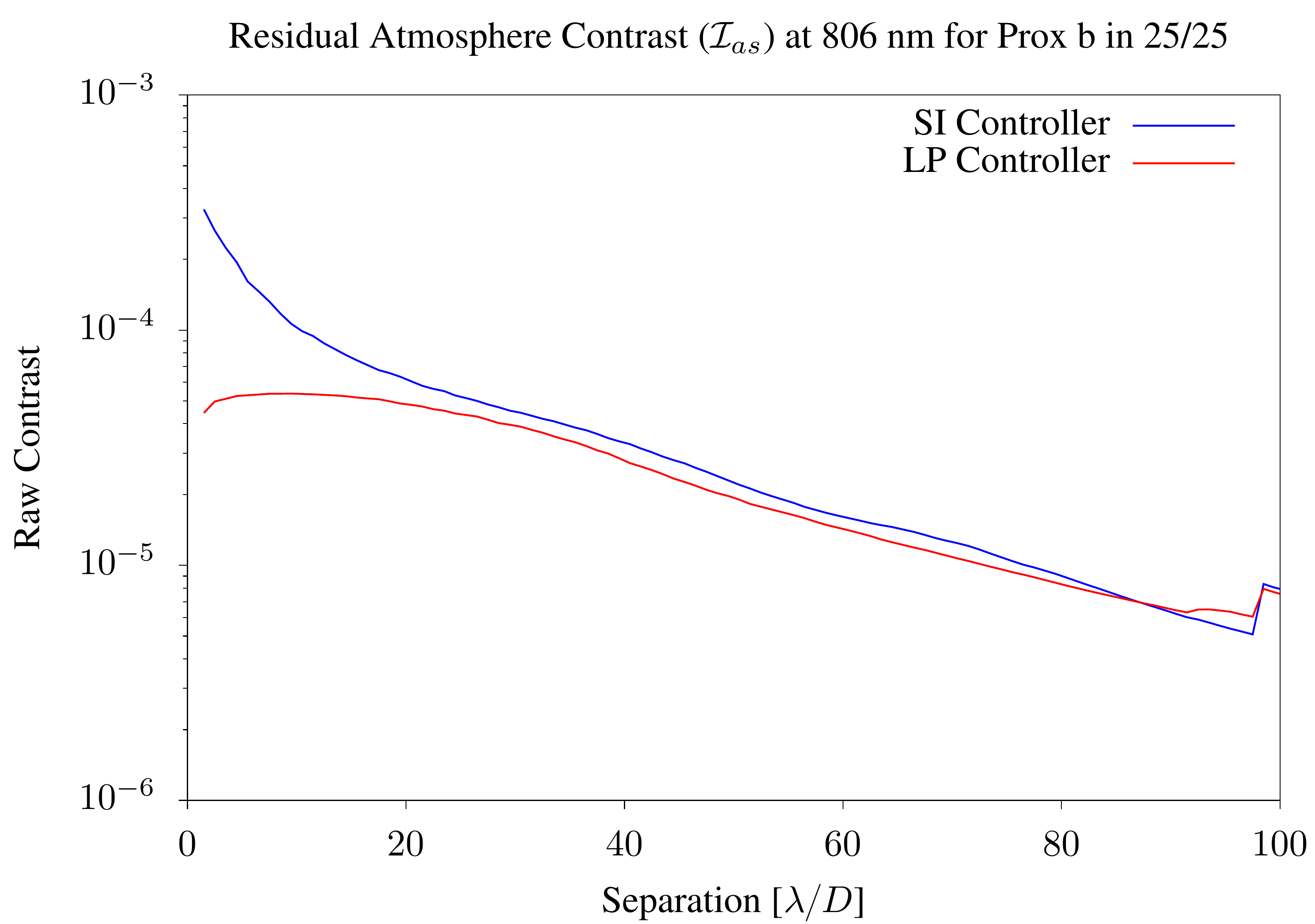}
\includegraphics[width=2.5in]{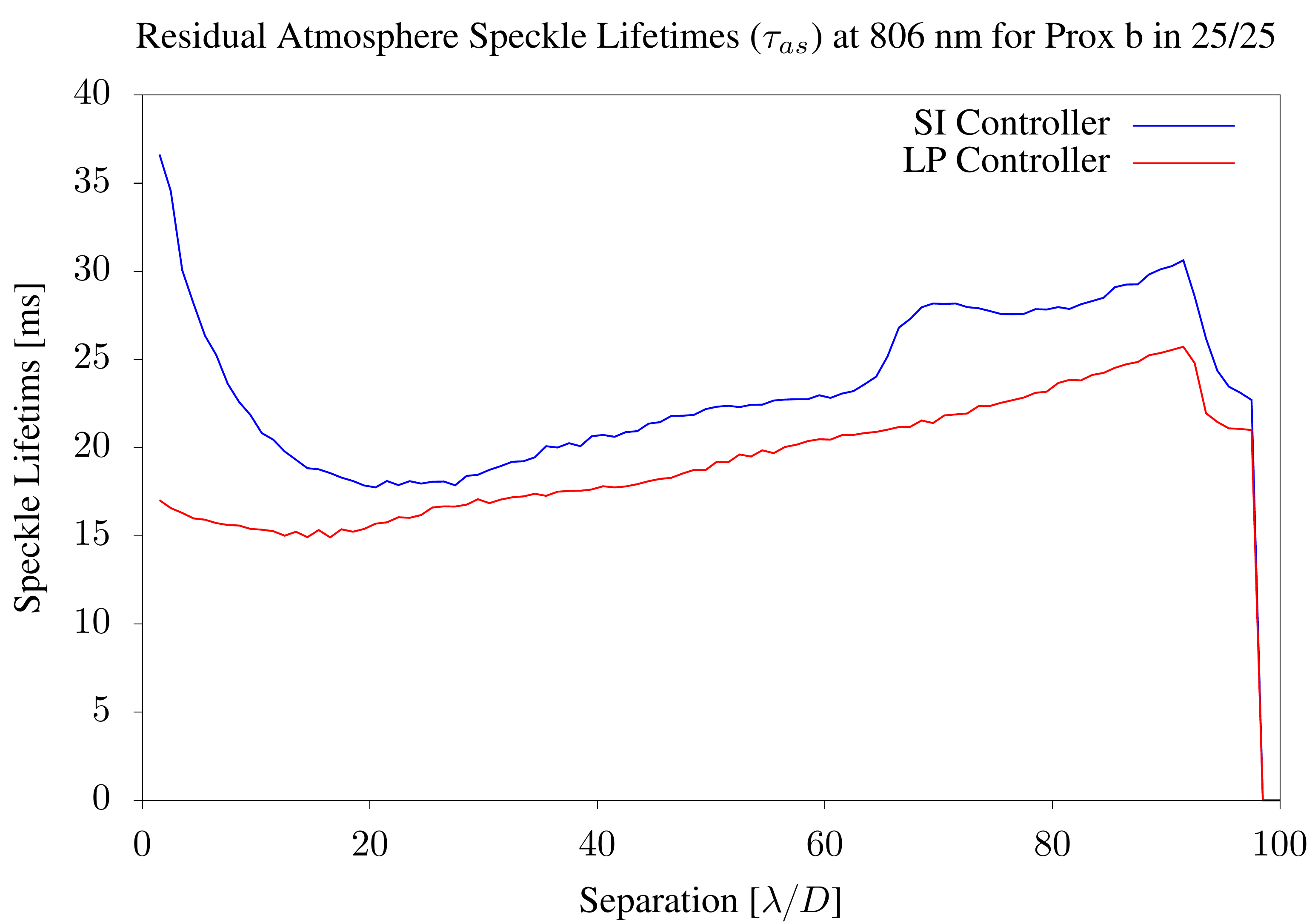}
\caption{Left: GMagAO-X residual atmospheric instantaneous contrast.  Right: GMagAO-X residual atmosphere speckle lifetime.  Blue is for the simple integrator controller, and red is for the ideal
linear predictor controller. \label{fig:contrast}}
\end{figure}

A dynamical model of the closed-loop performance was used to predict the performance of the high-order AO system.  Figure \ref{fig:strehl} shows the Strehl ratio vs. guide-star magnitude.
Figure \ref{fig:contrast} shows the residual instantaneous contrast due to residual atmospheric speckles, as well as the statistical speckle lifetimes.

\begin{figure}
\centering
\includegraphics[width=6.5in]{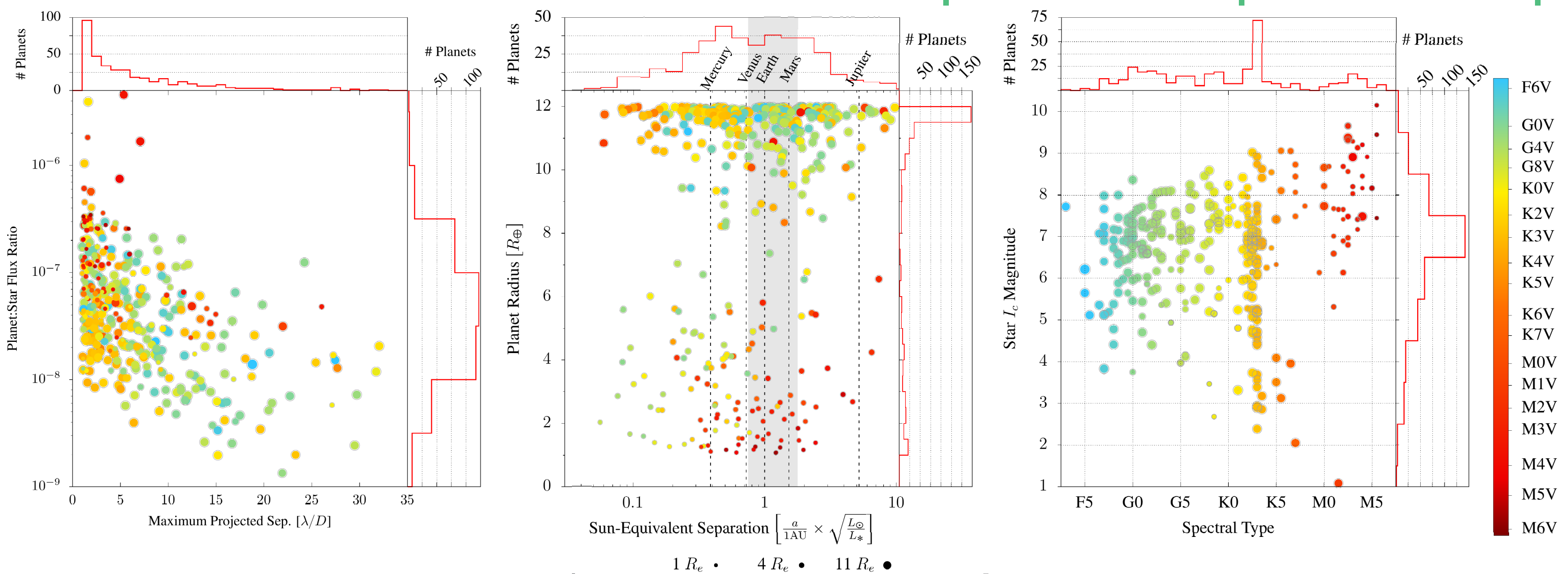}
\caption{Characteristics of the currently known exoplanets which GMagAO-X will characterize.
Left: Planet:Star flux ratio vs separation.  Middle: Planet radius vs. luminosity normalized semi-major axis.
Right: Guide-star magnitude vs spectral type for the planet host stars.\label{fig:yield}}
\end{figure}

The results of the performance analysis were used to assess the reflected light imaging exoplanet yield.  We base this analysis on the known exoplanets rather than a projected population model.  We emphasize this point: the planets under consideration are known to exist.  We assessed each planet based on its guide-star magnitude and the resultant predicted GMagAO-X performance.  For each planet we estimate its radius based on the RV minimum mass using a mass-to-radius relationship calibrated from measured exoplanets.  Geometric albedos were based on a suite of models including Earthshine, Venus, and published models for EGPs.  The results are shown in Figure \ref{fig:yield}.  GMagAO-X will be capable of characterizing the atmospheres of up to over 200 currently known exoplanets, for which we currently know only a minimum mass.

\section{CONCLUSION}

GMagAO-X is the ExAO-fed coronagraph planned for first-light of the GMT.  It represents the earliest opportunity of the ELT era to begin search nearby terrestrial planets for life.  GMagAO-X has passed preliminary design review, and is preparing to begin the final design phase.  The scientific potential of GMagAO-X highlights how crucial the US-ELT program is to the future of U.S. Astronomy and motivates a positive decision to move forward with construction of the GMT as rapidly as possible.

\acknowledgments 
The GMagAO-X conceptual and preliminary design would not have been possible without the support of the University of Arizona Space Institute.  We are also grateful for the support of an anonymous donor to Steward Observatory.

\bibliography{report} 
\bibliographystyle{spiebib} 

\end{document}